\documentclass[english,amssymb,amsmath,noshowpacs,prb,twocolumn]{revtex4-2}
\usepackage[english]{babel}
\usepackage[utf8]{inputenc}
\usepackage{amssymb,amsmath}

\usepackage[colorlinks,pdfpagelabels,pdfstartview = FitH,bookmarksopen = true,bookmarksnumbered = true,linkcolor = black,plainpages = false,hypertexnames = false,citecolor = black,urlcolor=black] {hyperref}
\usepackage{hyperref}
\usepackage{mathtools}

\begin{document}
	
	\title{Doping-induced non-Markovian interference causes \\ excitonic linewidth broadening in monolayer WSe\textsubscript{2}}
	
	\author{Florian Katsch}
	\author{Andreas Knorr}	
	\affiliation{Institut f\"ur Theoretische Physik, Nichtlineare Optik und Quantenelektronik, Technische Universit\"at Berlin, 10623 Berlin, Germany}
	
	\begin{abstract}		
		Strong Coulomb interactions in atomically thin semiconductors like monolayer WSe\textsubscript{2} induce not only tightly bound excitons, but also make their optical properties very sensible to doping.
		By utilizing a microscopic theory based on the excitonic Heisenberg equations of motion, we systematically determine the influence of doping on the excitonic linewidth, lineshift, and oscillator strength.
		We calculate trion resonances and demonstrate that the Coulomb coupling of excitons to the trionic continuum generates a non-Markovian interference, which, due to a time retardation, builds up a phase responsible for asymmetric exciton line shapes and increased excitonic linewidths.
		Our calculated doping dependence of exciton and trion linewidths, lineshifts, and oscillator strengths explains recent experiments.
		The gained insights provide the microscopic origin of the optical fingerprint of doped atomically thin semiconductors.
		\\
	\end{abstract}
	
	\maketitle

\textit{Introduction}. Excitons are quasi-particle excitations built up from Coulomb-bound electron-hole pairs in semiconductors which share few optical properties with atomic two-level systems \cite{allen1987optical}.
	Instead, most excitonic properties are determined by many-body phenomena without counterparts in atomic systems, including exciton-exciton, exciton-electron, and exciton-phonon interactions \cite{haug2009quantum}.
	Ideal prerequisites for the identification of novel many-exciton effects are provided by atomically thin semiconductors, in which two-dimensional confinement and reduced environmental screening cause exceptionally strong Coulomb interactions \cite{wang2018colloquium,gies2021atomically}.
	A monolayer of the transition metal dichalcogenide WSe\textsubscript{2} constitutes a typical atomically thin semiconductor where optical exciton resonances form several hundreds of meV below the band edge \cite{ramasubramaniam2012large,berkelbach2013theory,he2014tightly,trushin2016optical,trushin2018model,deilmann2019finite}.
	The steering of those tightly bound excitonic quantum states could pioneer novel optoelectronic and quantum technological applications \cite{butler2013progress,jariwala2014emerging,mueller2018exciton}.
	But strong Coulomb interactions in monolayer WSe\textsubscript{2} also imply a sensible doping dependence of the excitonic properties, confirmed in gate-voltage-dependent experiments	\cite{jones2013optical,shepard2017trion,roch2018quantum,roch2019spin,smolenski2019interaction,goldstein2020ground,wagner2020autoionization,xiao2020many,liu2020gate}.
	Hence, already a weak unintentional doping \cite{mak2013tightly,chernikov2015electrical,chen2018coulomb,li2018revealing,liu2019gate}, originating from impurities, vacancies, or substrate effects, can corrupt the preparation of pure excitonic quantum states due to a pronounced doping-induced linewidth broadening of excitons \cite{goldstein2020ground,wagner2020autoionization,xiao2020many,wu2021enhancement} and trion resonances \cite{mak2013tightly,mouri2013tunable,courtade2017charged,florian2018dielectric}.
	Although technological applications of excitonic quantum states in atomically thin semiconductors require an understanding of doping-induced many-body phenomena without counterparts in atomic systems, a microscopic theory is still missing.
	Here, we introduce a microscopic many-body theory based on coupled Maxwell's and excitonic Bloch equations to describe the doping-dependent excitonic properties of atomically thin semiconductors.
	We focus on the regime of low doping densities, captured by a linear doping dependence and explicitly consider the coupling of excitons to trions and their corresponding scattering continua \cite{esser2001theory}.
	While trions are bound solutions of the Schrödinger equation, describing the interaction of a virtual exciton with the Fermi sea of dopants \cite{sidler2017fermi}, the scattering continua represent unbound solutions \cite{companion1}.
	We diagonalize the excitonic and trionic Schrödinger equations assuming an effective mass approximation, fully thermalized doping distributions, and an ansatz to separate the relative- and center-of-mass-motion.
	Trions form spectrally localized excitations below the exciton resonances.
	In contrast, the trionic scattering continuum starts above the exciton energy and produces a doping-dependent dephasing and asymmetric exciton linewidth broadening due to the non-Markovian interference of excitons with the trionic continuum.
	Incorporating excitons, trions and the entire trionic continuum, obtained by an exact diagonalization approach, our doping-dependent calculations of exciton and trion oscillator strengths, energy separations, and linewidth broadenings agree with a recent experiment from Wagner \textit{et al.} \cite{wagner2020autoionization}.
	Due to the lack of exactly calculated scattering continua, previous theories only explained specific aspects of the experimentally observed features like doping-dependent excitonic and trionic oscillator strengths \cite{glazov2020optical,rana2020many,efimkin2021electron}.
	Predicted exciton-trion energy separations already spanned from values comparable to experiments \cite{sidler2017fermi,chang2019many,fey2020theory,glazov2020optical,efimkin2021electron} to overestimates \cite{rana2020many}.
	However, no description could quantitatively explain the interplay of measured exciton-trion energy separation and simultaneously occurring exciton linewidth broadening, which was underestimated in previous theories \cite{chang2019many,rana2020many,carbone2020microscopic,efimkin2021electron} compared to experiments \cite{goldstein2020ground,wagner2020autoionization,xiao2020many}.
	Our description manifests that the interplay and description of simultaneous energy shifts and linewidth broadenings requires a consistent many-body treatment of not only excitons and trions but also the associated scattering continua obtained by exact diagonalization.

	\begin{figure}
		\centering
		\includegraphics[width=1\columnwidth]{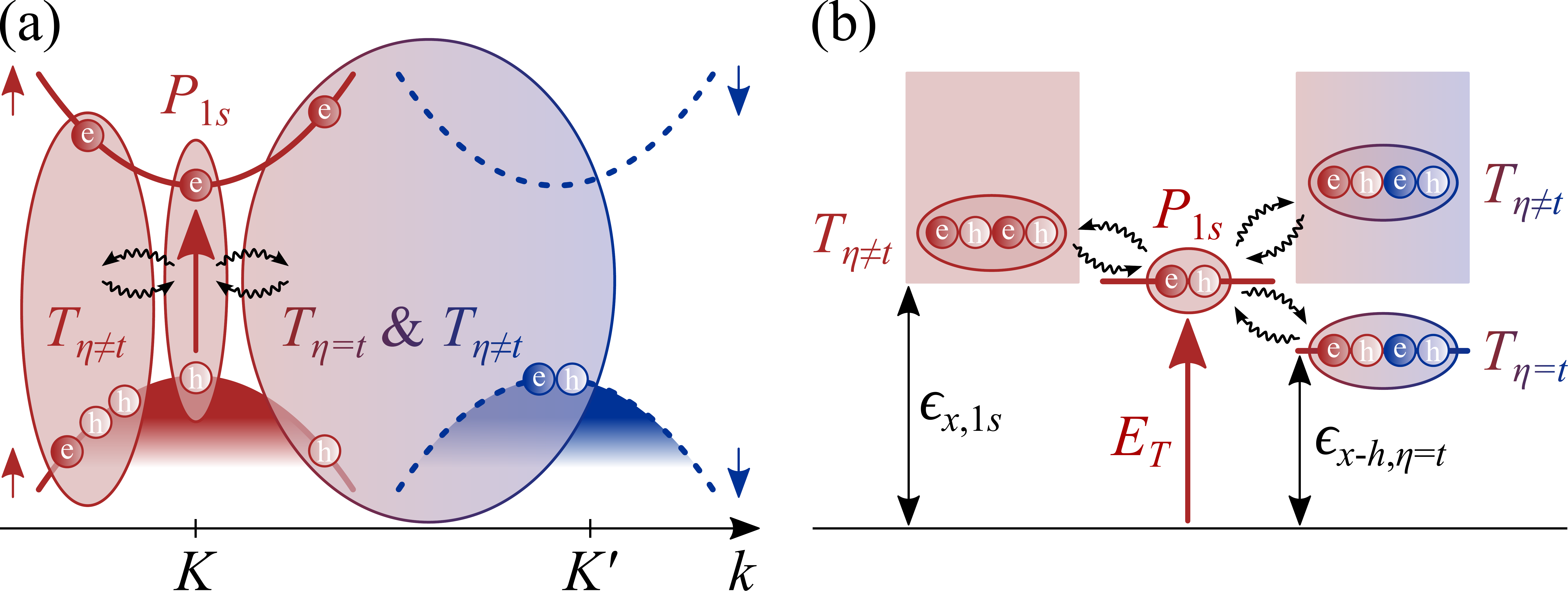}
		\caption{
			(a)~Illustration of optically excited excitons $P_{1s}$, as bound electron-hole pairs, which interact with trions $T_{\eta=t}$ and the scattering continua $T_{\eta\neq t}$ in a reduced band structure at the $K$ and $K'$ high-symmetry points.
			(b)~Illustration of the coupled states.
			Solid lines depict the spectrally localized excitations $P_{1s}$ and $T_{\eta=t}$ with energies $\epsilon_{x,1s}$ and $\epsilon_{x\text{-}h,\eta=t}$, respectively.
			Shaded areas represent the scattering continua $T_{\eta\neq t}$ starting at $\epsilon_{x,1s}$.
			Excitons $P_{1s}$ are excited by the light field $E_T$.
		}
		\label{Bild:Energie-Schema}
	\end{figure}

\textit{Description of excitons}. The exciton energies $\epsilon_{x,\lambda_1}$ and wave functions~$\varphi_{\lambda_1,\textbf{\textit{k}}}$ of bound electron-hole pairs, illustrated in Fig.~\ref{Bild:Energie-Schema}(a), are obtained by solving the Wannier equation \cite{kira2006many} for monolayer WSe\textsubscript{2} encapsulated in hexagonal BN in an effective mass approximation \cite{parametersWSe2}.
	The index $\lambda_1$ comprises the high-symmetry point ($K$, $K'$) and spin ($\uparrow$, $\downarrow$) of the electron and hole together with the exciton state ($1s$, $2s$, $2p^\pm$, ...).
	Since we subsequently focus on the hole doping regime, the low energy band structure can be described by the energetically highest valence bands and energetically lowest conduction bands with equal spins around the $K$ and $K'$ points, respectively.
	After the Wannier equation is solved, the electron-hole transitions are characterized by excitonic dipole interband transitions $P_{\lambda_1}$ \cite{katsch2018theory}, which act as a direct source in Maxwell's equations \cite{knorr1996theory,jahnke1997linear}.
	The dynamics of the optically addressable exciton transitions $P_{\lambda_1}$, drawn as a red solid line in Fig.~\ref{Bild:Energie-Schema}(b), reads \cite{companion1}:
	\begin{eqnarray}
	& & \left(\hbar\partial_t + \Gamma_{x} - i \epsilon_{x,\lambda_1} \right) P_{\lambda_1} \notag \\
	& & = -i \sum_{\textbf{\textit{k}}} d_{\lambda_1,\textbf{\textit{k}}} (1-f_{\textbf{\textit{k}}}) E_T(t)
	+ i \sum_{\lambda_2,\textbf{\textit{k}}} {W}_{\lambda_1,\lambda_2,\textbf{\textit{k}}} \ f_{\textbf{\textit{k}}} \ P_{\lambda_2}
	\notag \\
	& & \hspace{3.8mm} + i \sum_{\eta,\textbf{\textit{k}}} \hat{W}_{\lambda_1,\eta,\textbf{\textit{k}}} \ f_{\textbf{\textit{k}}} \ T_{\eta} .
	\label{eq:eom-x}
	\end{eqnarray}
	Equation~\eqref{eq:eom-x} represents excitonic oscillations with energy $\epsilon_{x,\lambda_1}$ which are damped by a phonon-mediated dephasing $\Gamma_{x}$ \cite{selig2016excitonic,brem2019intrinsic}.
	The first term on the right-hand side of Eq.~\eqref{eq:eom-x}, including the interband transition dipole element $d_{\lambda_1,\textbf{\textit{k}}}$, represents the optical source and Pauli blocking due to a doping density $f_{\textbf{\textit{k}}}$.
	$E_T(t)$ denotes the light field which introduces a radiative dephasing after self-consistently solving Eq.~\eqref{eq:eom-x} and Maxwell's equations \cite{knorr1996theory,jahnke1997linear}.
	The second contribution to the right-hand side of Eq.~\eqref{eq:eom-x}, connected with the Coulomb matrix element ${W}_{\lambda_1,\lambda_2,\textbf{\textit{k}}}$, describes a Coulomb-induced reduction of the quasi-particle band gap and exciton binding energy depending on the doping density $f_{\textbf{\textit{k}}}$.
	The last contribution to Eq.~\eqref{eq:eom-x}, involving the Coulomb matrix element $\hat{W}_{\lambda_1,\eta,\textbf{\textit{k}}}$, couples excitons $P_{\lambda_1}$ to trions $T_{\eta=t}$ and trionic continuum transitions $T_{\eta\neq t}$.
	The index $\eta$ includes the high-symmetry points and spins of comprised electrons and holes along with the state index indicating bound (trion) and unbound (continuum) states, see Fig.~\ref{Bild:Energie-Schema}.
	Bound and unbound states are provided by the Schrödinger equation of a virtual exciton interacting with the Fermi sea of dopants \cite{sidler2017fermi} described by a fully thermalized Fermi distribution, where, similar to the Wannier equation, the relative- and center-of-mass-motion wave functions were separated \cite{companion1}.
	Repulsive Coulomb interactions between a virtual exciton and the Fermi sea in the same band preclude intravalley trions ($\eta=t$) and only provide intravalley continuum states ($\eta\neq t$).
	The intravalley continuum starts above the exciton energy $\epsilon_{x\text{-}h,\eta\neq t}\geq \epsilon_{x,1s}$ and is drawn as a red shaded area in Fig.~\ref{Bild:Energie-Schema}(b).
	In contrast, attractive Coulomb interactions between a virtual exciton and the Fermi sea in different bands provide bound states described as intervalley trions in addition to an intervalley continuum.
	The intervalley trion appears energetically below the exciton resonance $\epsilon_{x\text{-}h,\eta=t}<\epsilon_{x,1s}$ and is illustrated as a solid line in Fig.~\ref{Bild:Energie-Schema}(b).
	The intervalley continuum sets in at the exciton energy $\epsilon_{x\text{-}h,\eta\neq t}\geq \epsilon_{x,1s}$ and is depicted as a shaded area in Fig.~\ref{Bild:Energie-Schema}(b).
	The dynamics of trions $T_{\eta=t}$ and scattering continua $T_{\eta\neq t}$ reads \cite{companion1}:
	\begin{equation}
	\left[ \hbar \partial_t + \Gamma_{\eta} - i \epsilon_{x\text{-}h,\eta} \right] T_{\eta} = i \sum_{\lambda_1,\textbf{\textit{k}}} \tilde{W}_{\eta,\lambda_1,\textbf{\textit{k}}} \ P_{\lambda_1} \ f_{\textbf{\textit{k}}} . \label{eq:eom-trion}
	\end{equation}
	Equation~\eqref{eq:eom-trion} describes oscillations with energy $\epsilon_{x\text{-}h,\eta}$ which are damped by a phonon-mediated dephasing $\Gamma_{\eta}$ \cite{triondeph}.
	The oscillations are driven by an exciton $P_{\lambda_1}$ and the doping density $f_{\textbf{\textit{k}}}$ proportional to the Coulomb matrix element $\tilde{W}_{\eta,\lambda_1,\textbf{\textit{k}}}$.
	Inserting the formal integration of the trionic continuum $T_{\eta\neq t}$, Eq.~\eqref{eq:eom-trion}, into the exciton dynamics $P_{\lambda_1}$, Eq.~\eqref{eq:eom-x}, induces a time retardation and develops a scattering-induced phase resulting in a non-Markovian interference of Coulomb-coupled excitons and trionic continua.
	Note that Eq.~\eqref{eq:eom-trion} implies the truncation to linear doping densities where filling factors $(1-f_{\textbf{\textit{k}}})$ in Eq.~\eqref{eq:eom-trion} are neglected.
	This approximation is validated for low enough doping densities $N_h$ ensuring that the Fermi distribution $f_{\textbf{\textit{k}}}$ is much less extended in momentum space compared to the trion wave function represented by the trion Bohr radius at, i.e., $N_h a_t^2 \ll 1 $ \cite{esser2001theory}.

\textit{Absorption spectra}. Self-consistently solving the dynamics of coupled exciton, trion, and scattering continua, Eqs.~\eqref{eq:eom-x} and \eqref{eq:eom-trion}, with Maxwell's equations for a normal incidence geometry \cite{knorr1996theory} in frequency ($\omega$) domain determines the doping-dependent absorption spectrum $\alpha_{1s}(\omega)$ at the $1s$ exciton resonance $P_{1s}$ \cite{companion1}:
	\begin{equation}
		\alpha_{1s}(\omega) \hspace{-0.6mm} =  \hspace{-0.6mm}
		\frac{2\Gamma_{r}\left[\Gamma_{x} +\Gamma_{x\text{-}h}(\omega)\right]}{\left[\epsilon_{x,1s} + \Delta_{x\text{-}h}(\omega) -\hbar\omega\right]^2+\left[\Gamma_{x} + \Gamma_{r}+\Gamma_{x\text{-}h}(\omega)\right]^2}
		 . \label{eq:refl}
	\end{equation}
	%
	$\Gamma_{r}$ is the radiative dephasing originating from the optical source represented by the first term on the right-hand side of Eq.~\eqref{eq:eom-x} including Pauli blocking.
	The frequency-dependent dephasing $\Gamma_{x\text{-}h}(\omega)$ and renormalization $\Delta_{x\text{-}h}(\omega)$ of the exciton energy $\epsilon_{x,1s}$ stem from Coulomb interactions, see also Ref.~\cite{companion1}:
	The second term on the right-hand side of Eq.~\eqref{eq:eom-x} induces a frequency-independent contribution to the lineshift $\Delta_{x\text{-}h}(\omega)$ \cite{steinhoff2014influence,erben2018excitation}. 
	The frequency-dependence of the lineshift $\Delta_{x\text{-}h}(\omega)$ and dephasing $\Gamma_{x\text{-}h}(\omega)$ exclusively originate from the coupling of excitons $P_{1s}$ to trions $T_{\eta=t}$ and the scattering continua $T_{\eta\neq t}$, described by the last contribution to Eq.~\eqref{eq:eom-x}.
	Trions $T_{\eta=t}$ contribute an exciton-trion level repulsion to $\Delta_{x\text{-}h}(\omega)$ and the trion resonance to $\Gamma_{x\text{-}h}(\omega)$.
	The scattering continua $T_{\eta\neq t}$ renormalize the lineshift $\Delta_{x\text{-}h}(\omega)$ and introduce a frequency-dependent dephasing to $\Gamma_{x\text{-}h}(\omega)$.
	The latter originate from the non-Markovian interference of Coulomb-coupled excitons $P_{1s}$ and trionic continua $T_{\eta\neq t}$ obtained by inserting the formal integration of Eq.~\eqref{eq:eom-trion} into Eq.~\eqref{eq:eom-x}.

	\begin{figure}
		\centering
		\includegraphics[width=1.0\columnwidth]{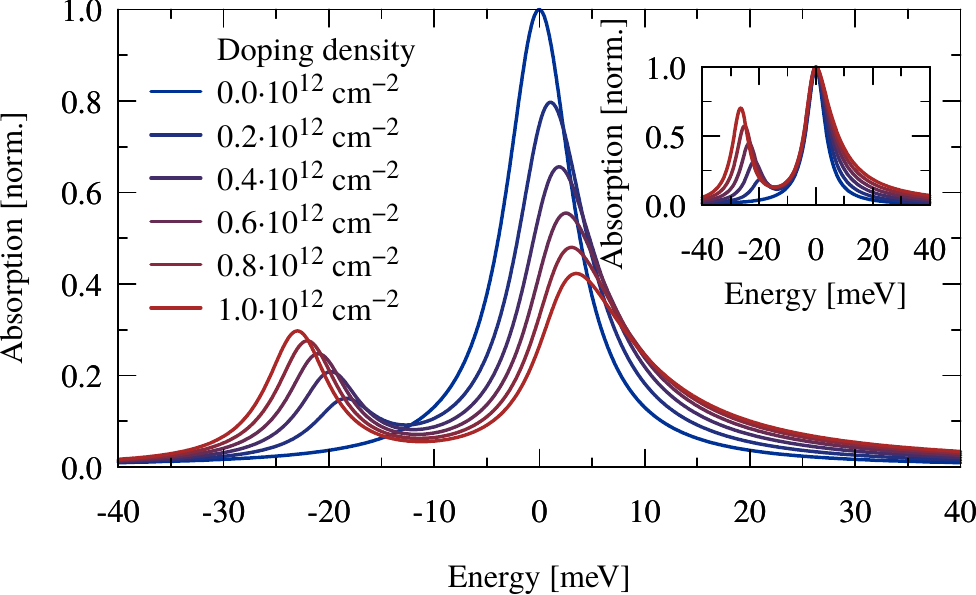}
		\caption{
			Absorption spectra for monolayer WSe\textsubscript{2} encapsulated in hexagonal BN near the $1s$ exciton resonance at 5~K.
			The doping density of free holes in the valence bands rises from $0~$cm$^{-2}$ (charge neutrality) to $10^{12}~$cm$^{-2}$.
			The origin of the energy scale coincides with the charge neutral exciton energy.
			The spectra are normalized with respect to the undoped sample.
			Inset: Normalized and shifted absorption spectra.
		}
		\label{Bild-Exzitonen}
	\end{figure}

\begin{figure}
		\centering
		\includegraphics[width=1.0\columnwidth]{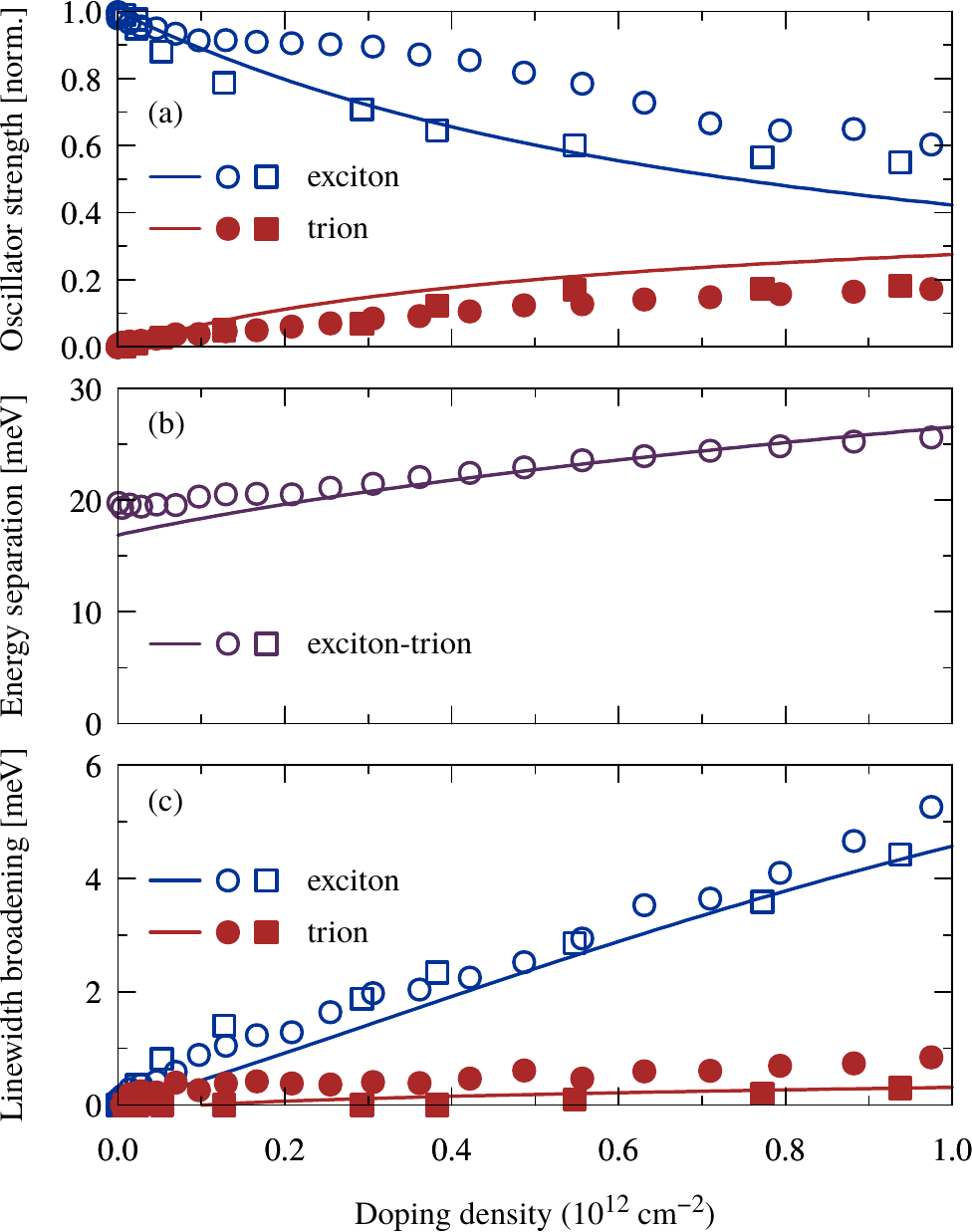}
		\caption{
			Doping-density-dependent (a)~oscillator strengths of $1s$ excitons and trions, (b)~their energy separation, and (c)~the exciton and trion linewidth broadenings (full width at half maximum) subtracted from the homogeneous linewidth at charge neutrality.
			The results are obtained for monolayer WSe\textsubscript{2} encapsulated in hexagonal BN at 5~K from theory (solid lines) and experiment from Ref.~\cite{wagner2020autoionization} (circles, squares).
		}
		\label{Bild-Comp}
	\end{figure}

	Figure~\ref{Bild-Exzitonen} shows the calculated absorption spectra for monolayer WSe\textsubscript{2} encapsulated by hexagonal BN at 5~K for different doping densities.
	With rising doping, (a)~the exciton oscillator strength decreases while a trion resonance emerges, (b)~the exciton shifts toward higher energies and the trion toward lower energies, and (c)~the exciton linewidth broadens.
	We now separately discuss the different observations in Fig.~\ref{Bild-Comp} and compare our calculations to experiments from Wagner \textit{et al.} \cite{wagner2020autoionization}.

(a)~\textit{Oscillator strengths}. The blue and red solid lines in Fig.~\ref{Bild-Comp}(a) compare the calculated exciton and trion oscillator strengths.
	With increasing doping density, the exciton oscillator strength decreases and the trion oscillator strength increases \cite{note_trion_osc}.
	Our calculations are compared with measurements from two different monolayer WSe\textsubscript{2} samples from Ref.~\cite{wagner2020autoionization}, plotted as circles and squares in Fig.~\ref{Bild-Comp}(a).
	Overall, we find a reasonable agreement between theory (solid lines) and experiment (circles, squares).
	The decreasing exciton oscillator strength with rising doping primarily stems from a Coulomb-induced redistribution of oscillator strength from excitons $P_{1s}$ to trions $T_{\eta = t}$ and the scattering continua $T_{\eta\neq t}$ related to $\Gamma_{x\text{-}h}(\omega)$ in Eq.~\eqref{eq:refl}.
	Although the oscillator strength of the trionic continuum $T_{\eta\neq t}$ cannot be measured directly, sidebands appear on the high energy side of the exciton resonance, responsible for the linewidth broadening discussed later.
	Pauli blocking accounts for only five percent of the reduced exciton oscillator strength at the considered low doping densities.

(b)~\textit{Energy renormalizations}. The purple solid line in Fig.~\ref{Bild-Comp}(b) shows the calculated energy separation between exciton and trion resonances at different doping densities.
	At negligible doping, the energy separation approaches the theoretical trion binding energy of 17~meV close to the experimental value of 19~meV from Ref.~\cite{wagner2020autoionization} plotted as circles in Fig.~\ref{Bild-Comp}(b).
	As the doping increases, the exciton-trion energy separation rises and we find an agreement between theory (solid line) and experiment (circles) from Ref.~\cite{wagner2020autoionization} in Fig.~\ref{Bild-Comp}(b).
	The increasing energy separation originates from Coulomb interactions of excitons $P_{1s}$, trions $T_{\eta = t}$, and scattering continua $T_{\eta\neq t}$ comprised in the renormalization $\Delta_{x\text{-}h}(\omega)$ in Eq.~\eqref{eq:refl}.
	In $\Delta_{x\text{-}h}(\omega)$, Coulomb interactions reduce the quasi-particle band gap which is overcompensated by a decreased exciton binding energy resulting in a blue shift of the exciton resonance of several meV \cite{Ataei2021competitive}.
	The exciton-trion level repulsion increases this excitonic blue shift, while the repulsion with the trionic continuum reduces it. All effects contribute to an increase of the total exciton-trion energy separation of a few meV.
	Although experiments of various materials on different substrates observed trion red shifts \cite{,mak2013tightly,jones2013optical,chernikov2015electrical,goldstein2020ground,wagner2020autoionization,xiao2020many}, some experiments found trion blue shifts at high doping densities, where trions dominate the absorption and excitons have vanishing oscillator strengths \cite{van2019probing,liu2020gate}. Here, the interplay of nonlinear doping effects and dynamical screening might lead to a different behavior compared to the low doping regime.
	Weaker trion red shifts for electron doping \cite{wagner2020autoionization} might be due to energetically low lying spin-split conduction bands and minima at the $\Lambda$ high-symmetry points \cite{kormanyos2015k}.

(c)~\textit{Linewidth broadening}. Figure~\ref{Bild-Comp}(c) depicts the exciton and trion linewidth broadening, presented as full widths at half maxima without the radiative and phonon-mediated linewidth contributions $2(\Gamma_{r}+\Gamma_{x})$ at charge neutrality.
	The theoretical exciton linewidth broadening is plotted as a blue solid line in Fig.~\ref{Bild-Comp}(c) and almost linearly grows by 4.7~meV per doping density of 10$^{12}$~cm$^{-2}$ in agreement with measurements from Ref.~\cite{wagner2020autoionization} plotted as blue circles and squares \cite{SiO2_substrat}.
	The trion linewidth broadening is largely doping-independent in theory (red solid line) and experiment (red circles, squares).
	A linearly increasing exciton linewidth with rising doping density and a nearly doping-independent trion linewidth broadening was also experimentally observed in other atomically thin semiconductors at weak doping densities \cite{goldstein2020ground,xiao2012coupled,wu2021enhancement}.
	\begin{figure}
		\centering
		\includegraphics[width=1.0\columnwidth]{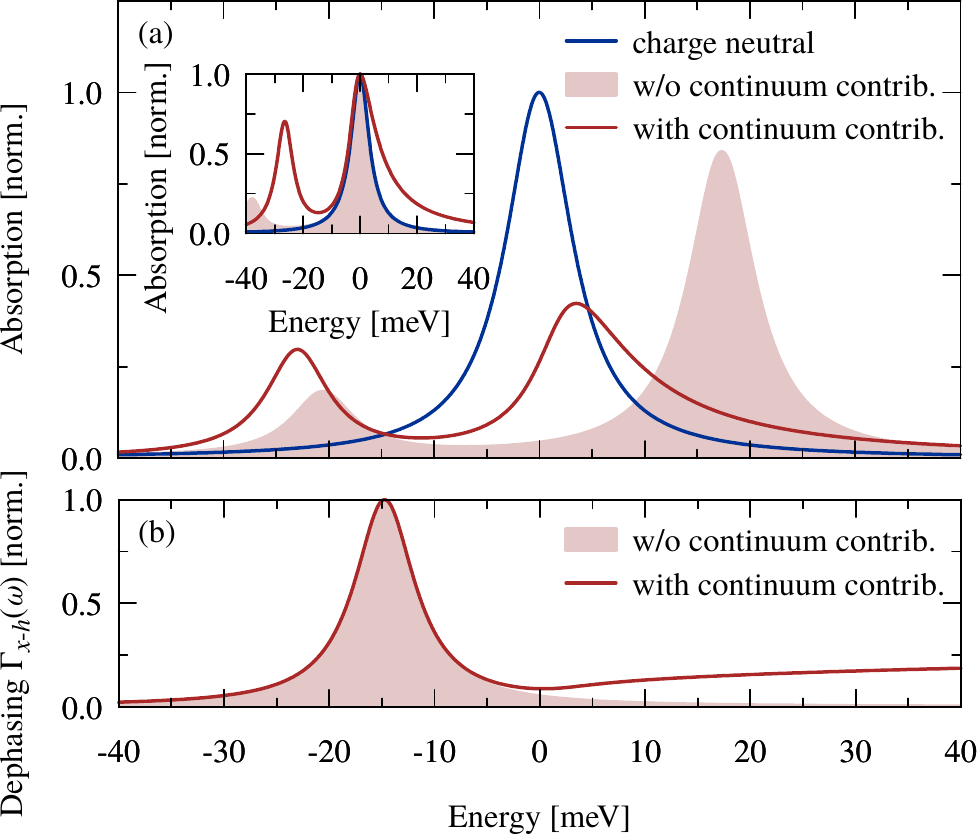}
		\caption{
			(a)~Absorption spectra for monolayer WSe\textsubscript{2} encapsulated in hexagonal BN near the $1s$ exciton resonance at 5~K.
			Shown is the absorption at charge neutrality and doping densities of free holes in the valence bands of $10^{12}~$cm$^{-2}$ with and without the trionic continuum.
			The origin of the energy scale coincides with the charge neutral exciton energy and the spectra are normalized with respect to charge neutrality.
			Inset: Normalized and shifted absorption.
			(b)~Normalized scattering-induced dephasing $\Gamma_{x\text{-}h}(\omega)$ at a doping density of $10^{12}~$cm$^{-2}$ with and without the trionic continuum.
		}
		\label{Bild-Kont}
	\end{figure}
	To disentangle the microscopic origin of the exciton linewidth broadening, Fig.~\ref{Bild-Kont}(a) displays the influence of the scattering continua.
	The blue solid line in Fig.~\ref{Bild-Kont}(a) represents the charge neutral absorption and the red shaded area is the absorption at a doping density of $10^{12}~$cm$^{-2}$ without the coupling of excitons $P_{1s}$ to the scattering continua $T_{\eta\neq t}$.
	Although the pure exciton-trion interaction can qualitatively explain a reduced exciton oscillator strength and renormalized exciton and trion resonance energies, the exciton linewidth is unchanged compared to no doping, see inset of Fig.~\ref{Bild-Kont}(a).
	The absorption including the full Coulomb interaction between excitons $P_{1s}$, trions $T_{\eta= t}$, and scattering continua $T_{\eta\neq t}$ is plotted as a red solid line in Fig.~\ref{Bild-Kont}(a).
	Here, the level repulsion between exciton and energetically higher continuum states leads to a red shift compared to the case without continuum contributions \cite{blueshift}.
	Moreover, a pronounced shoulder emerges on the high energy side of the exciton resonance, which induces an asymmetric exciton line shape increasing the exciton linewidth.
	By simultaneously solving the coupled Maxwell's and excitonic Bloch equations, we verified that the asymmetric exciton line shape is not due to the light-matter interaction.
	While previous calculations predicted pronounced sidebands at much larger doping densities \cite{rana2020many}, we predict sidebands at much weaker doping and enhanced linewidth broadenings in agreement with experiments \cite{wagner2020autoionization}.
	Although absorption measurements found sidebands on the low \cite{van2017marrying} and high energy sides \cite{goldstein2020ground,jadczak2021probing} of the exciton resonance, our consistent description of trionic many-particle interactions implies high-energy sidebands \cite{sideband}.
	The high energy sideband originates from the frequency dependence of the scattering-induced dephasing $\Gamma_{x\text{-}h}(\omega)$ entering Eq.~\eqref{eq:refl} induced by the non-Markovian interference discussed above.
	Without the trionic continuum $T_{\eta\neq t}$ the dephasing $\Gamma_{x\text{-}h}(\omega)$ is described by the trion state $T_{\eta= t}$ plotted as a red shaded Lorentzian in Fig.~\ref{Bild-Kont}(b) \cite{reduced}.
	The full dephasing $\Gamma_{x\text{-}h}(\omega)$ is shown as a red solid line in Fig.~\ref{Bild-Kont}(b).
	The nearly constant contribution to $\Gamma_{x\text{-}h}(\omega)$ above the exciton energy stems from the trionic continuum $T_{\eta\neq t}$, which induces a high energy sideband and broadens the exciton linewidth due to a redistribution of oscillator strength from excitons $P_{1s}$ to the trionic continuum $T_{\eta\neq t}$.
	At almost vanishing doping the trionic continuum $T_{\eta\neq t}$ sets in at the exciton energy, but with rising doping density the onset of the trionic continuum $T_{\eta\neq t}$, similar to the exciton energy, shifts towards higher energies due to Coulomb renormalizations.
	The doping-induced asymmetric linewidth broadening resembles the non-Markovian coupling of excitons to the phonon continuum \cite{rudin2002effects,christiansen2017phonon,shree2018observation,lengers2020theory,funk2021spectral} or the coupling among excitons and the exciton-exciton scattering continua \cite{katsch2020exciton,katsch2020optical}.
	But suppressed phonon populations at cryogenic temperatures and weak exciton densities at low excitation powers yield vanishing phonon-induced and exciton-scattering-induced sidebands.
	Hence, doping-induced high energy sidebands of the exciton resonance naturally occur in doped semiconductors.
	Although dynamical screening effects can qualitatively explain the doping dependence of excitonic properties and asymmetric resonances, its impact is only prominent at considerable doping densities \cite{gao2016dynamical,van2017marrying,van2019probing,scharf2019dynamical} but negligible at lower doping densities where the trionic continuum dominates the line shape.
	However, at high doping densities frequency-dependent Coulomb scattering beyond the static approximation \cite{glazov2018breakdown} and nonlinear doping contributions need to be included in the theory, which will likely result in a nonlinear dependence of the exciton linewidth broadening on the doping density.
	In contrast, a low energy exciton side band appears for continua below the exciton energy, similar to asymmetric $\pi$ excitons in graphene which couple to the energetically lower Dirac continuum of electronic states \cite{yang2009excitonic,mak2011seeing,chae2011excitonic}.
\textit{Conclusion}. We demonstrated that the non-Markovian interference of Coulomb coupling between excitons and trionic scattering continua is responsible for a frequency-dependent dephasing, which leads to an asymmetric exciton linewidth broadening.
	This asymmetric broadening of exciton resonances leaves a pronounced fingerprint in the absorption spectra besides trion resonances.
	We directly demonstrated that our calculated doping-dependent exciton and trion linewidth broadenings, oscillator strengths, and their energy separation agree with experiments from Wagner \textit{et al.} \cite{wagner2020autoionization}.
	In conclusion, our results open a pathway to theoretically understand the influence of doping on the excitonic properties of organic or inorganic semiconductors with strong Coulomb interactions on a microscopic footing.
	\textit{Acknowledgments}. We thank Dominik Christiansen and Malte Selig (TU Berlin) for many stimulating discussions.
	Additionally, we thank Koloman Wagner (University of Regensburg), Alexey Chernikov (TU Dresden), Mikhail M. Glazov (Ioffe Institute) and all involved coworkers for valuable exchange and for allowing us to reproduce their experimental data.
	We gratefully acknowledge support from the Deutsche Forschungsgemeinschaft through Project No.~420760124 (\mbox{KN 427/11-1}).


\begin{thebibliography}{78}
\bibitem [{tri()}]{triondeph}%
	\BibitemOpen
	\href@noop {} {}\bibinfo {note} {{For simplicity the phonon-mediated
			dephasing $\Gamma_{\eta}$ of trions and scattering continua is adjusted to
			the excitonic one $\Gamma_{x}$.}}\BibitemShut {Stop}%
\bibitem [{not()}]{note_trion_osc}%
	\BibitemOpen
	\href@noop {} {}\bibinfo {note} {{The trion oscillator strength is obtained
			after subtracting the background signal of the exciton
			resonance.}}\BibitemShut {Stop}%
\bibitem [{SiO()}]{SiO2_substrat}%
	\BibitemOpen
	\href@noop {} {}\bibinfo {note} {Placing monolayer WSe\textsubscript{2} on a
		SiO\textsubscript{2} substrate, instead of encapsulation by hexagonal BN,
		leads to a slower decrease of the exciton oscillator strength with rising
		doping density, an increased trion binding energy of 22~meV, a weakly
		increased exciton-trion energy separation, and a weaker exciton linewidth
		broadening which almost linearly grows by 4.3~meV per doping density of
		$10^{12}~$cm$^{-2}$.}\BibitemShut {Stop}%
\bibitem [{blu()}]{blueshift}%
	\BibitemOpen
	\href@noop {} {}\bibinfo {note} {{Note that the absorption still exhibits a
			total reduced blue shift compared to charge neutrality since the combined
			influence of reduced quasi-particle band gap, decreased exciton binding
			energy, and exciton-trion level repulsion is not
			overcompensated.}}\BibitemShut {Stop}%
\bibitem [{red()}]{reduced}%
	\BibitemOpen
	\href@noop {} {}\bibinfo {note} {{The decreased trion energy of 15~meV in
			Fig.~\ref{Bild-Kont}(b) compared to the binding energy of 17~meV at charge
			neutrality is due to Coulomb-induced renormalizations.}}\BibitemShut {Stop}%
\end{thebibliography}
\end{document}